\documentclass[journal]{IEEEtran}
\usepackage{amsmath}
\usepackage{float}
\usepackage{graphicx}
\usepackage{multirow}
\usepackage{amsfonts,amssymb}
\usepackage{bm}
\usepackage{cite}  
\usepackage{color}
\newtheorem{lemma}{Lemma}

\ifCLASSINFOpdf
\else
\fi
\begin{document}
%
\title{ Performance Analysis on Machine Learning-Based Channel Estimation}


\author{Kai~Mei,~Jun~Liu,~Xiaochen~Zhang,~Nandana~Rajatheva~and~Jibo Wei
\thanks{(c) 2021 IEEE. Personal use is permitted, but republication/redistribution requires IEEE permission. See https://www.ieee.org/publications/rights/index.html for more information. Citation information for this article : DOI 10.1109/TCOMM.2021.3083597.}
\thanks{Kai Mei, Jun Liu, Xiaochen Zhang, and Jibo Wei are with the
College of Electronic Science and Technology, National University of Defense
Technology, Changsha 410073, China (E-mail: {meikai11, liujun15, zhangxiaochen14, wjbhw}@nudt.edu.cn).}
\thanks{Nandana Rajatheva is with Center for Wireless Communications, University of Oulu, Oulu 90570, Finland (E-mail: nandana.rajatheva@oulu.fi).}
}


%



\IEEEtitleabstractindextext{%
\begin{abstract}

Recently, machine learning-based channel estimation has attracted much attention. The performance of machine learning-based estimation has been validated by simulation experiments. However, little attention has been paid to the theoretical performance analysis. In this paper, we investigate the mean square error (MSE) performance of machine learning-based estimation. Hypothesis testing is employed to analyze its MSE upper bound. Furthermore, we build a statistical model for hypothesis testing, which holds when the linear learning module with a low input dimension is used in machine learning-based channel estimation, and derive a clear analytical relation between the size of the training data and performance. Then, we simulate the machine learning-based channel estimation in orthogonal frequency division multiplexing (OFDM) systems to verify our analysis results. Finally, the design considerations for the situation where only limited training data is available are discussed. In this situation, our analysis results can be applied to assess the performance and support the design of machine learning-based channel estimation.

\end{abstract}

}

\maketitle

\IEEEdisplaynontitleabstractindextext

%
\IEEEpeerreviewmaketitle

\section{Introduction}
%
%
%
%

\IEEEPARstart{I}{n} wireless communication systems, the transmitted signals are corrupted by many detrimental factors, such as multipath propagation, and mobility, etc \cite{7501500,MATZ20111}. To recover the transmitted data accurately, channel estimation is an essential module in the coherent receiver. A portion of the transmitted signals are known at the receiver and used for the channel estimation, which are called pilot signals \cite{851324}. Among the pilot-aided channel estimation methods, least-squares (LS) estimation has the lowest complexity, which simply divides the received signal value by the pilot \cite{7890505,4267831}. However, its performance is sensitive to noise. To improve the performance of LS estimation, the correlation in time, frequency, and space domain can be exploited in linear minimum mean square error (LMMSE) estimation, where the correlated LS estimates are used to enhance an estimate of interest \cite{6847111,1033009}. Although LMMSE estimation has the optimal estimation performance \cite{504981}, the computational complexity is much higher than LS estimation and it requires the knowledge of second-order channel statistics. Moreover, LMMSE estimation draws on the condition that the channel is linear and stationary. When the channel is non-linear or non-stationary, LMMSE estimation suffers from performance degradation and the closed-form expression of the optimal estimation, i.e., minimum mean square error (MMSE) estimation, turns out to be intractable \cite{8272484,KangDeep}.

While there were early works on machine learning (ML)-based channel estimation \cite{zhou2003channel,zhang2007mimo,omri2010channel}, the recent re-emergence of machine learning has motivated the use of neural networks for channel estimation again \cite{8272484,KangDeep,8663458,8798971,8933411,9067011,8933050,9048929,8672767,8847452,8640815,he2018deep,8715649}. In ML-based channel estimation, the LS estimates are fed into a neural network, and then the neural network yields the enhanced channel estimates. As a model-free approach, ML-based channel estimation merely needs a dataset labeled by true channel responses to optimize the parameters of the neural network. An effective estimator can still be learned under complex channel conditions. In \cite{8272484}, a non-stationary channel is considered, where the expression of the MMSE estimation cannot be calculated in closed form. The heuristic structure deduced under a given distribution of the second-order channel statistics is used as a blueprint to design the neural network. Through training the neural network, the estimator performs well under arbitrary channel models. Until now, ML-based channel estimation has been developed for wireless energy transfer (WET) systems \cite{KangDeep}, orthogonal frequency division multiplexing (OFDM) systems \cite{8663458}, multiple-input multiple-output (MIMO) \cite{8272484,8798971,8933411}, and massive MIMO \cite{9067011,8933050,9048929}. Deep learning (DL) techniques including fully connected deep neural network (FC-DNN) \cite{8672767}, recurrent neural network (RNN) \cite{8847452}, and convolution neural network (CNN) \cite{8640815,he2018deep} have been leveraged to enhance the channel estimation performance.

In the literature, the merits of ML-based channel estimation have been demonstrated by experiments. However, the theoretical analysis of ML-based channel estimation has been paid little attention. Specifically, it has been shown by simulations that ML-based channel estimation can approach the optimal estimation \cite{8272484}, i.e., MMSE estimation. The mean square error (MSE) difference between the two methods, however, has not been investigated analytically. In addition to performance, the requirements on the training dataset, which plays an important role in the ML-based channel estimation, have not been studied analytically as well. In this paper, we analyze the ML-based channel estimation from a theoretical perspective, and our main contributions are the following.
\begin{itemize}
\item We investigate the MSE difference between the ML-based channel estimation and the MMSE channel estimation employing hypothesis testing. The analysis shows that there is an upper bound of the MSE difference even if the inputs are not from the training dataset, which provides theoretical support for ML-based channel estimation.

\item We further specify the distribution functions in hypothesis testing, which holds for the situation where the learning module is linear with a low input dimension. Then, we define the scaled MSE difference as the performance measure of the ML-based channel estimation and derive the analytical relation between the training dataset size and performance. Moreover, we display the numerical result of this analytical relation and obtain the sufficient size of the training dataset.

\item We conduct computer simulations to evaluate the performance of ML-based channel estimation, where the linear structure and deep neural network (DNN) are employed. Through simulation experiments, the derived sufficient size of the training dataset is examined and our theoretical analysis is validated.

\item We discuss the design of ML-based channel estimation for the situation where only limited training data is available. Since the linear learning module may be employed in this situation, our analysis results can be used to predict the performance of ML-based channel estimation and help determine the input dimension of the learning module.

\end{itemize}

The rest of this paper is organized as follows. Channel models, the MMSE estimation, and the ML-based channel estimation are introduced in Section \ref{sec.CE}. The performance analysis on ML-based channel estimation is presented in Section \ref{sec.PA}. To validate our analysis, the performance of ML-based channel estimation employing the linear structure and DNN is evaluated through simulations in Section \ref{sec.EAR} and Section \ref{sec.CLM}, respectively. The design considerations of ML-based channel estimation with limited training data are discussed in Section \ref{sec.MLCS}. Finally, the paper is concluded in Section \ref{sec.Conl}.

Notation: We use boldface small letters and capital letters to denote vectors and matrices. $\mathbb{E}\left[  \cdot  \right]$, $\mathbb{D}\left[  \cdot  \right]$,  and ${\left\|  \cdot  \right\|_2}$ represent the expectation, the variance, and the Euclidean norm, respectively. The superscripts ${\left(  \cdot  \right)^ * }$, ${\left(  \cdot  \right)^{\rm{T}}  }$, ${\left(  \cdot  \right)^{\rm{H}}  }$, ${\left(  \cdot  \right)^{ - 1}}$ denote the conjugate of complex, the transpose, the Hermitian transpose of a complex vector or matrix, and the inversion, respectively. ${\mathcal{CN}\left( \cdot \right)}$ represents the complex Gaussian distribution, while ${\chi ^2}\left( {\kappa} \right)$ represents the chi-square distribution with dimension of $\kappa$. We denote the probability density function (PDF) and cumulative probability function (CDF) of ${\chi ^2}\left( {\kappa} \right)$ as ${p_{\chi _\kappa ^2}} ( \cdot ) $ and ${F_{\chi _\kappa ^2}( \cdot )}$, respectively.

%

\section{Channel Estimation}
\label{sec.CE}

\subsection{Channel Model and MMSE Estimation}
\label{sec.CM}

In pilot-assisted channel estimation, known pilot signals are transmitted and used for channel estimation. LS estimation is usually adopted to estimate the channel responses  as the initial estimates and then these initial estimates can be improved with many different methods \cite{4267831}.

The LS estimation simply divides the received signal values by the pilot values \cite{4267831}. We use ${{{\bf{\hat h}}}_{\rm{p}}}$ to represent a $N_{\rm p} \times 1$ vector that contains the LS estimates, where $N_{\rm p}$ is the number of pilot signals. ${{\hat {\bf h}}_{\rm{p}}}$ can be modeled as the superposition of actual channel responses and noise \cite{4267831,1033009}, i.e.,
\begin{equation}
{{{\bf{\hat h}}}_{\rm{p}}} = {{{\bf{h}}}} + {\bf n},
\end{equation}
where ${\bf n}$ is a white Gaussian noise vector with variance ${\sigma^2}$. ${\bf{h}}$ contains the true channel responses for pilot signals.

We denote $\hat {\bf h}_{\rm s}$ as a $N_{\rm p} \times 1$ vector containing the improved estimates based on ${{{\bf{\hat h}}}_{\rm{p}}}$ and use a function ${\bf f}(\cdot)$ to represent a certain channel estimation method \cite{8640815}, i.e.,

\begin{equation}
\label{equ.EstVec}
{\hat {\bf h}_{\rm s}} = {\bf f}\left( {{{\bf{\hat h}}}_{\rm{p}}} \right).
\end{equation}

We aim to analyze the performance of channel estimation in this paper and the performance analysis is usually focused on a single channel response, e.g., in \cite{1673071}. Therefore, we consider only one estimate from the final output of channel estimation. We define $\hat h_{\rm s}$ as an arbitrary element in $\hat {\bf h}_{\rm s}$ and investigate the performance of $\hat h_{\rm s}$. The index of $\hat h_{\rm s}$ is omitted for simplicity since the position of $\hat h_{\rm s}$ in $\hat {\bf h}_{\rm s}$ is not concerned. Then, we have

\begin{equation}
\label{equ.Est}
{{\hat h}_{\rm s}} = f\left( {{{\bf{\hat h}}}_{\rm{p}}} \right).
\end{equation}

Note that the performance analysis on $\hat h_{\rm s}$ can be used to assess a channel estimation method represented by (\ref{equ.EstVec}). Since $\hat h_{\rm s}$ is an element of the vector $\hat {\bf h}_{\rm s}$, the analysis can be focused on $\hat h_{\rm s}$ while investigating the estimation performance of $\hat {\bf h}_{\rm s}$ in (\ref{equ.EstVec}).

As a general representation, (\ref{equ.Est}) can describe a wide range of channel estimation methods. The design of a channel estimation method is to pursue a low MSE and we denote $f_{{\rm{opt}}} ( \cdot )$ as the one that has the minimal MSE, i.e., the MMSE estimation \cite{kay1993fundamentals}. The analytical expression of $ f_{{\rm{opt}}}\left( \cdot \right)$ depends on the statistical model of the channel \cite{8272484}. In this paper, we consider two types of channel models: the stationary channel model and the quasi-stationary channel model.

The stationary channel which is subject to complex Gaussian distribution is often assumed for conventional channel estimation methods \cite{701321}. Then, we have ${{{\bf{h}}}} \sim {\cal C}{\cal N}( {{\bf 0},{{\mathbf{R}}_{{{\bf{h}}}{{\bf{h}}}}}} )$ and the MMSE estimation can be expressed as \cite{4267831}

\begin{equation}
\label{equ.mmse}
f_{{\rm{opt}}}\left( {{{\bf{\hat h}}}_{\rm{p}}} \right) = {{\mathbf{r}}_{h_{\rm{s}}{{\bf{h}}}}}{\left( {{{\mathbf{R}}_{{{\bf{h}}}{{\bf{h}}}}} + {\sigma^2}{\mathbf{I}}} \right)^{ - 1}}{{{\bf{\hat h}}}_{\rm{p}}},
\end{equation}
where $h_{\rm{s}}$ is the actual channel response of $\hat h_{\rm s}$. ${{\mathbf{r}}_{h_{\rm{s}}{{\bf{h}}}}}$ denotes the correlation vector between $h_{\rm{s}}$ and ${{\bf{h}}}$, i.e., ${{\mathbf{r}}_{h_{\rm{s}}{{\bf{h}}}}} = \mathbb{E}[ {h_{\rm{s}}{{( {{{\bf{h}}}} )^{\text{H}}}}} ]$. ${\bf{I}}$ is an identity matrix.

Channel estimation for the quasi-stationary channel has been investigated in the recent literature, e.g., in \cite{8272484}. Under a quasi-stationary channel, ${{\mathbf{r}}_{h_{\rm{s}}{{\bf{h}}}}}$ and ${{\mathbf{R}}_{{{\bf{h}}}{{\bf{h}}}}}$ are not fixed but depend on the given parameters $\bm{\delta}$. $\bm{\delta}$ may describe, for example, angles of propagation paths and the parameters in $\bm{\delta}$ are assumed to be random variables \cite{8272484}. Then, ${{{\bf{h}}}}$ is assumed to be conditionally Gaussian distributed, i.e. ${{{\bf{h}}}}\left| {\bm{\delta}} \right. \sim {\cal C}{\cal N}\left( {0,{{\bf{R}}_{{\bf{hh}}\left| {\bm{\delta}} \right.}}} \right)$. Under this condition, the optimal estimation is given by \cite{8272484}
\begin{equation}
\label{equ.NonLOpt}
\begin{aligned}
{f_{{\rm{opt}}}}\left( {{\hat {\bf{h}}}_{\rm{p}}} \right) &= \mathbb{E}\left[ {{{\bf{h}}}\left| {{{\bf{\hat h}}}_{\rm{p}}} \right.} \right] \cr 
& = \mathbb{E}\left[ {\mathbb{E}\left[ {{{\bf{h}}}\left| {{{\bf{\hat h}}}_{\rm{p}}} \right.,{\bm{\delta}} } \right]\left| {{{\bf{\hat h}}}_{\rm{p}}} \right.} \right] \cr 
& = \mathbb{E}\left[ {{{\bf{f}}_{{\rm{opt}}\left| {\bm{\delta}} \right.}}\left( {{{\bf{\hat h}}}_{\rm{p}}} \right)\left| {{{\bf{\hat h}}}_{\rm{p}}} \right.} \right] ,
\end{aligned}
\end{equation}
where $$
{f_{{\rm{opt}}\left| {\bm{\delta}} \right.}}\left( {{{\bf{\hat h}}}_{\rm{p}}} \right)={{\bf{r}}_{h_{\rm{s}}{{\bf{h}}}\left| {\bm{\delta}} \right.}}{\left( {{{\bf{R}}_{{{\bf{h}}}{{\bf{h}}}\left| {\bm{\delta}} \right.}} + {\sigma^2}{\bf{I}}} \right)^{ - 1}}{{{\bf{\hat h}}}_{\rm{p}}}.
$$ It is difficult to derive the exact expression of (\ref{equ.NonLOpt}).

The analytical expression of $ f_{{\rm{opt}}}\left( \cdot \right)$ or MMSE estimation depends on the channel model. Therefore, when we develop a channel estimation algorithm based on MMSE estimation, we should investigate the channel model. Furthermore, when the channel model is complicated, the exact expression of MMSE estimation is hard to obtain. As a result, the MMSE channel estimation may sometimes be practically infeasible.

\subsection{Machine Learning-based Channel Estimation}

\begin{figure}
\begin{centering}
\includegraphics[scale=0.7]{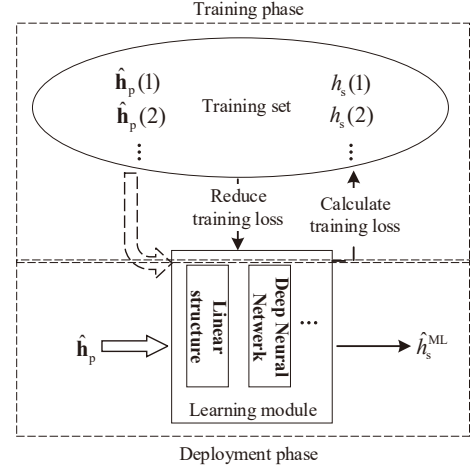}

\end{centering}
\caption{Sketch diagram of machine learning based estimation.}
\label{fig.MLEstStruct}
\end{figure}

Channel estimation can be realized in a quite different manner by leveraging machine learning. The procedure of machine learning-based channel estimation is illustrated in Fig. \ref{fig.MLEstStruct} \cite{8272484}. The key component is the learning module, which is employed to approximate the function in (\ref{equ.Est}). Convolution neural network (CNN) \cite{8640815}, recurrent neural network (RNN) \cite{8847452}, the linear structure and etc. can be used as the learning module. The linear structure directly connects the output with the input and is the simplest learning module, which can only fit a linear function.

There are two phases in machine learning-based channel estimation including the training phase and the deployment phase. In the training phase, the parameters of the learning module are optimized through reducing a loss function over a data set $ {\cal T}$. To be specific, the data set $ {\cal T}$ can be represented as ${\cal T} = \{ ({\bf{\hat h}}_{\rm{p}}(1),{h_{\rm{s}}}(1))...({\bf{\hat h}}_{\rm{p}}(m),{h_{\rm{s}}}(m))...({\bf{\hat h}}_{\rm{p}}(M),{h_{\rm{s}}}(M))\} $, where $({\bf{\hat h}}_{\rm{p}}(m),{h_{\rm{s}}}(m))$ denotes the $ m {\rm th}$ pair of training data in $\cal T$ and ${h_{\rm{s}}}(m)$ is the label for the input ${\bf{\hat h}}_{\rm{p}}(m)$. Note that we omit the index $m$ for simplicity if it is not needed. The loss function is defined as the square error of estimation, i.e., ${{\cal{L}}( {f( {{\bf{\hat h}}_{\rm{p}}} ),{h_{\rm{s}}}} )}= { | {f( {{\bf{\hat h}}_{\rm{p}}} )-{h_{\rm{s}}}}  |^2}$. In addition, we define ${\cal L}_{\cal T }$ as the average loss function over the data set ${\cal T }$, i.e., 
\begin{equation}
\label{eq.Trainingloss}
{\cal L}_{\cal T } = {1 \over M}\sum\limits_m {\left | {f\left( {{\bf{\hat h}}_{\rm{p}}\left( m \right)} \right) - {h_{\rm{s}}}\left( m \right)} \right |^2}. 
\end{equation}
We call ${\cal L}_{\cal T }$ the training loss in the following. Through minimizing ${\cal L}_{\cal T }$, the learning module can be trained to approximate a function that achieves good estimation performance. In the deployment phase, the initial estimates ${{\bf{\hat h}}_{\rm{p}}}$ input the learning module and then the learning module produces the estimate of $h_{\rm{s}}$.

The machine learning-based channel estimation does not heavily rely on the channel model \cite{8272484} and merely needs a dataset for training. This is because the learning module of machine learning-based channel estimation can be simply regarded as a black box, which can directly perform channel estimation after its parameters are optimized through training. It is not required to derive the explicit expression of channel estimation, unlike conventional methods. As a result, when a complicated channel condition is considered, an effective estimator can still be learned in a data-driven manner.

However, the theoretical analysis on the performance of machine learning-based channel estimation lacks in the recent literature. If the sample size is infinitely large and the training loss approaches the expected square error, i.e., MSE, MMSE estimation can be learned by training since the estimator that minimizes the training loss will be the one that minimizes MSE. However, in practice, the sample size is a finite value and the training loss is merely the sampled value of MSE. Then, the training procedure can only guarantee that the learned estimator minimizes the square error of training data. If new data inputs the estimator, the MSE performance of the output is unpredictable and only experimental evaluation has been provided for the performance of ML-based channel estimation. Therefore, we analyze the MSE performance of the ML-based channel estimation in this paper.

\section{Performance Analysis on Machine Learning-based Channel Estimation}
\label{sec.PA}

\subsection{Performance Analysis Using Hypothesis Testing}
\label{SVML}

The MSE of channel estimation represented by $ f \left(  \cdot  \right)$ is also the expectation of loss function, as can be seen from the equation below.
\begin{equation}
\label{equ.Eloss}
\begin{aligned}
{{\cal{L}}_{\rm{E}}} &={\mathbb E}\left[ {{\cal{L}}\left( {f \left( {{\bf{\hat h}}_{\rm{p}}} \right),{h_{\rm{s}}}} \right)} \right] \cr
&= {\mathbb E} \left[ {\left | {f\left( {{\bf{\hat h}}_{\rm{p}}} \right)-{h_{\rm{s}}}} \right |^2} \right].
\end{aligned}
\end{equation}
It is difficult to obtain the joint probability density function (PDF) of $f ( {{\bf{\hat h}}_{\rm{p}}} )$ and ${h_{\rm{s}}}$, which depends on many channel statistics, and thus the MSE of $ f \left(  \cdot  \right)$ is intractable.

We denote $ f_*\left(  \cdot  \right)$ as the learned function in ML-based channel estimation. The learned estimation is normally not the MMSE estimation $f_{{\rm{opt}}} ( \cdot )$ as mentioned above, and there is a certain loss in the MSE performance compared with $f_{{\rm{opt}}} ( \cdot )$. We denote the MSEs of ${f_{{\rm{opt}}}\left(  \cdot  \right)}$ and ${f_*\left(  \cdot  \right)}$ as  ${{\cal{L}}_{{\rm{E1}}}}$ and ${{\cal{L}}_{{\rm{E2}}}}$, respectively. We use ${\Delta _{{{\cal{L}}_E}}}$ to represent the MSE difference between ${f_{{\rm{opt}}}\left(  \cdot  \right)}$ and ${f_*\left(  \cdot  \right)}$, i.e., ${\Delta _{{{\cal{L}}_E}}} = {{\cal{L}}_{{\rm{E2}}}}-{{\cal{L}}_{{\rm{E1}}}} $. Notice that the MSE of $ f_*\left(  \cdot  \right)$ is difficult to calculate and one may concern more about the MSE difference ${\Delta _{{{\cal{L}}_E}}}$ than the exact MSE ${{\cal{L}}_{{\rm{E2}}}}$. Since the MSE difference ${\Delta _{{{\cal{L}}_E}}}$ can show whether the ML-based channel estimation approaches the optimal performance, it can indicate the learning performance more clearly than the exact value of MSE ${{\cal{L}}_{{\rm{E2}}}}$. We investigate the MSE difference ${\Delta _{{{\cal{L}}_E}}}$ in this paper.

We employ hypothesis testing to analyze ${\Delta _{{\cal{L}}_E}}$ \cite{lehmann2006testing}. Define ${\Delta _{{\cal{L}}_E}}  \ge  \Delta _{{\cal{L}}_E}^0$ as hypothesis $H_0$ and ${\Delta _{{\cal{L}}_E}}  <  \Delta _{{\cal{L}}_E}^0$ as hypothesis $H_1$. Set the confidence level as $1- {\varepsilon _0}$. Then, if the probability of the observed event that happens under hypothesis $H_0$ is lower than ${\varepsilon _0}$, i.e., $P (  {H_0}  ) \le {\varepsilon _0}$, we can accept $H_1$ and it is derived that the upper bound of the MSE difference $\Delta _{{\cal{L}}_E}$ is $\Delta _{{\cal{L}}_E}^0$. More specifically, we believe that the MSE difference between the MMSE estimation $f_{{\rm{opt}}} ( \cdot )$ and the learned estimation ${f_*\left(  \cdot  \right)}$ is no more than $ \Delta _{{\cal{L}}_E}^0$ at a confidence level of $1- {\varepsilon _0}$. Since whether the condition $P (  {H_0}  ) \le {\varepsilon _0}$ holds is unknown , further analysis on $P (  {H_0}  )$ is needed.

Denote ${\xi _1}$ as the training loss of ${f_{{\rm{opt}}}\left(  \cdot  \right)}$, i.e., $${\xi _1} = {1 \over M}\sum\limits_m {{{\left| {{f_{{\rm{opt}}}}\left( {{\bf{\hat h}}_{\rm{p}}\left( m \right)} \right) - {h_{\rm{s}}}\left( m \right)} \right|}^2}}. $$ Denote ${\xi _2}$ as the training loss of ${f_*\left(  \cdot  \right)}$, i.e., $${\xi _2} = {1 \over M}\sum\limits_m {{{\left| {{f_*}\left( {{\bf{\hat h}}_{\rm{p}}\left( m \right)} \right) - {h_{\rm{s}}}\left( m \right)} \right|}^2}} .$$ The learned estimation has the minimal training loss, i.e., ${\xi _1} \ge {\xi _2}$. We denote the probability that ${\xi _1} \ge {\xi _2}$ as $\varepsilon $. Note that $P (  {H_0}  )$ is the probability of ${\xi _1} \ge {\xi _2}$ under hypothesis $H_0$. We have $P\left(  {H_0}  \right) = \varepsilon$ when ${{\Delta _{{\cal{L}}_E}}  \ge  \Delta _{{\cal{L}}_E}^0}$. To simplify the expression of $\varepsilon$, we need the following assumption.

\emph{Assumption 1.} ${\xi _1}$ is independent of ${\xi _2}$, i.e., $p\left( {{\xi _1},{\xi _2}} \right) = {p_1}\left( \xi_1 \right){p_2}\left( \xi_2 \right)$, where ${p_1}\left( \xi_1 \right)$ and ${p_2}\left( \xi_2 \right)$ are the PDFs of ${\xi _1}$ and ${\xi _2}$, respectively.

If Assumption 1 does not hold, e.g., if $f_{\rm opt}(\cdot) = f_*(\cdot)$ and ${\xi _1}$ is highly correlated with ${\xi _2}$, the actual value of $P\left(  {H_0}  \right)$ is lower than its calculated value $\varepsilon$. To illustrate, when $f_*(\cdot)$ is close to $f_{\rm opt}(\cdot)$, which may violate the independence between ${\xi _1}$ and ${\xi _2}$, the MSE difference between $f_*(\cdot)$ and $f_{\rm opt}(\cdot)$ is reduced, which contributes to $H_1$. Then, the actual value $P\left(  {H_0}  \right)$ decreases and thus is lower than the calculated value $\varepsilon$. In the case, we can still accept $H_1$ at the same confidence level. Since the confidence level can be regarded as the lower bound of $P\left( {H_1} \right)$, the actual probability of $H_1$ can be higher than the confidence level. In other words, the hypothesis testing results derived under Assumption 1 apply for the situation where Assumption 1 does not hold.

Under Assumption 1, $\varepsilon $ can be expressed as
\begin{equation}
\label{equ.possib1}
\begin{aligned}
   \varepsilon & = \int_0^\infty  {\int_0^{{x_1}} {p\left( {{x_1},{x_2}} \right)} } d{x_2}d{x_1}  \cr 
  &  = \int_0^\infty  {{p_1}\left( {{x_1}} \right)\int_0^{{x_1}} {{p_2}\left( {{x_2}} \right)} } d{x_2}d{x_1}  \cr 
  &  = \int_0^\infty  {{p_1}\left( {{x_1}} \right){F_2}\left( {{x_1}} \right)} d{x_1},
\end{aligned}
\end{equation}
where ${F_2}\left( x \right)$ is the cumulative probability function (CDF) of ${\xi _2}$, i.e., ${F_2}\left( x \right) = \int_{ - \infty }^x {{p_2}\left( z \right)} dz$. 

\begin{figure}
\begin{centering}
\includegraphics[scale=0.5]{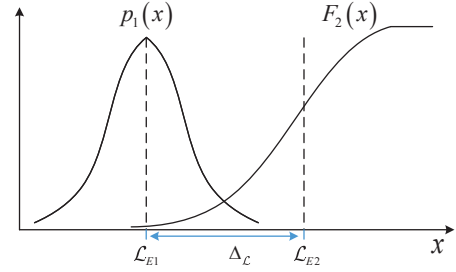}

\end{centering}
\caption{Sketch diagram for PDF of ${\xi_1}$ and CDF of ${\xi_2}$.}
\label{fig1}
\end{figure}

The value of $\varepsilon $ is dependent on ${\Delta _{{{\cal{L}}_E}}}$. Fig. \ref{fig1} displays an example of shapes for ${p_1}\left( x \right)$ and ${F_2}\left( x \right)$. With the increasing of ${\Delta _{{{\cal{L}}_E}}}$, the high value region of ${p_1}\left( x \right)$ will further move to the near zero region of ${F_2}\left( x \right)$. According to (\ref{equ.possib1}), when the multiplication of ${p_1}\left( x \right)$ and ${F_2}\left( x \right)$ is near zero, $\varepsilon $ will be rather small. Therefore, we can infer that $\varepsilon $ is negatively correlated to ${\Delta _{{{\cal{L}}_E}}}$.

We assume that the value of $\varepsilon$ is $\varepsilon_0$ when ${{\Delta _{{\cal L}_E}}} = \Delta _{{\cal L}_E}^0$. Since $P\left( {H_0} \right) = \varepsilon$ for ${{\Delta _{{\cal L}_E}}} \ge \Delta _{{\cal L}_E}^0$ and $\varepsilon $ is negatively correlated to ${\Delta _{{{\cal{L}}_E}}}$, $P\left( {H_0} \right)$ has the maximum value when ${\Delta _{{\cal L}_E}}$ achieves its minimum value $\Delta _{{\cal L}_E}^0$. As $\varepsilon = \varepsilon_0$ when ${{\Delta _{{\cal L}_E}}} = \Delta _{{\cal L}_E}^0$, the maximum value of $P\left( {H_0} \right)$ is $\varepsilon_0$. Therefore, we have $P\left( {H_0} \right) \le {\varepsilon _0}$. It demonstrates the condition $P\left( {H_0} \right) \le {\varepsilon _0}$ in above hypothesis testing and thus proves that compared with the optimal channel estimation, the MSE loss of machine learning-based channel estimation is less than $\Delta _{{\cal L}_E}^0$ at a confidence level of $1- {\varepsilon _0}$.

The above analysis actually demonstrates the validity of ML-based channel estimation, which is the main challenge when applying machine learning techniques. This is typically examined by experiments, while we verify it from a theoretical perspective. The analysis shows the MSE upper bound of ML-based channel estimation and the MSE is actually the expected loss function, as mentioned above. The expected loss function describes the performance of output for unseen data \cite{goodfellow2016deep} and thus indicates the applicability of machine learning when the input data is not from the training set.

From Fig. \ref{fig1}, the effect of the sample size on the performance of ML-based channel estimation can be explained as well. With the increasing of the sample size, the variances of $\xi_1$ and $\xi_2$ will be reduced, and then ${p_1}\left( x \right)$ will become narrow and ${F_2}\left( x \right)$ will increase more sharply. As a result, the overlapping part between ${p_1}\left( x \right)$ and ${F_2}\left( x \right)$ will be reduced and the value of $\varepsilon $ will be smaller. For the confidence level $1-\varepsilon_0 $, the corresponding upper bound $\Delta _{{\cal L}_E}^0$ decreases. Therefore, the MSE of the learned estimation ${{\cal{L}}_{{\rm{E2}}}}$ will be closer to the minimal MSE ${{\cal{L}}_{{\rm{E1}}}}$.

\subsection{Analytical Relation between Sample Size and Performance}
\label{Sec.ARSSP}

To derive the analytical relation between the sample size and $\Delta _{{\cal L}_E}^0$, the knowledge of the distributions of the training losses ${\xi _1}$ and ${\xi _2}$ are required. We make the following assumptions to specify a distribution for the training loss.

\emph{Assumption 2.} Output error is subject to complex Gaussian distribution, i.e., $( {f( {{\bf{\hat h}}_{\rm{p}}} ) - {h_s}} ) \sim {\cal C}{\cal N}\left( {0,{{\cal{L}}_{\rm{E}}}} \right)$. ${{\cal{L}}_{\rm{E}}}$ is actually the MSE of the estimation represented by ${f( {{\bf{\hat h}}_{\rm{p}}} )}$.

\emph{Assumption 3.} Output errors are independent, i.e., $({f ( {{\bf{\hat h}}_{\rm{p}}( m_1 )} ) - {h_s}( m_1 )})$ and $({f ( {{\bf{\hat h}}_{\rm{p}}( m_2 )} ) - {h_s}( m_2 )})$ are independent when ${m_1} \ne {m_2}$.

Assumption 2 holds when the simplest learning module is used, i.e., the linear structure. ${h_{\rm{s}}}$ and ${{\bf{\hat h}}_{\rm{p}}}$ are Gaussian random variables as mentioned in Section \ref{sec.CM}. The output error is the linear combination of those variables when the function $f(\cdot)$ is a linear one. Thus, the output error is Gaussian as well \cite{kay1993fundamentals}. Since the means of ${h_{\rm{s}}}$ and ${{\bf{\hat h}}_{\rm{p}}}$ are all zeros, the mean of the output error is also zero. In addition, since ${\mathbb{E}}[ {f( {{{{\bf{\hat h}}}_{\rm{p}}}} ) - {h_{\rm{s}}}} ] = 0$, the variance of the output error equals ${\mathbb{E}} [ {{{ | {f ( {{{{\bf{\hat h}}}_{\rm{p}}}}) - {h_{\rm{s}}}} |^2}}}]$, i.e., the MSE ${{\cal L}_E}$ for the channel estimation function $f(\cdot)$. Hence, we have $( {f( {{\bf{\hat h}}_{\rm{p}}} ) - {h_s}} ) \sim {\cal C}{\cal N}\left( {0,{{\cal{L}}_{\rm{E}}}} \right)$.

Assumption 3 approximately holds when the output errors are highly random \cite{draper1998applied}. To satisfy this condition, independent training data should be provided and the number of parameters in the learning module should be small. The independence of training data can be guaranteed by properly generating the inputs and labels \cite{8272484}. However, the number of parameters in the learning module is related to the chosen learning module and the use of a complex learning module may violate Assumption 3. As an illustration, if a complex learning module with a huge number of parameters is used, overfitting may happen \cite{goodfellow2016deep}. Then, the output errors may all be zeros and thus highly correlated.

In summary, Assumption 2 and Assumption 3 approximately hold when the learning module is approximately linear and the input dimension is low since a low input dimension can reduce the number of parameters in the learning module.

\begin{lemma}  
Given assumptions 2 and 3, the normalized training loss $2M{\xi }/{{\cal{L}}_{\rm{E}}}$ is subject to chi-square distribution ${\chi ^2}\left( {2M} \right)$, where ${\xi }$ is the training loss given in (\ref{eq.Trainingloss}).

Proof. If assumption 2 holds, we have ${{| {f ( {{\bf{\hat h}}_{\rm{p}}} )-{h_{\rm{s}}}}|^2}} = a^2+b^2$, where $a$ and $b$ are both Gaussian, i.e., $a \sim {\cal N}\left( {0,{{\cal{L}}_{\rm{E}}}/2} \right)$ and $b \sim {\cal N}\left( {0,{{\cal{L}}_{\rm{E}}}/2} \right)$. Then, $2{{| {f ( {{\bf{\hat h}}_{\rm{p}}} )-{h_{\rm{s}}} }|^2}/{{\cal{L}}_{\rm{E}}}}$ can be represented as the superposition of two normalized Gaussian variables.

If assumption 3 holds, the normalized training loss $2M{\xi }/{{\cal{L}}_{\rm{E}}}$ can be regarded as the superposition of the squares of $2M$ independent nomalized Gaussian variables, which is the formal description of the chi-square distribution ${\chi ^2}\left( {2M} \right)$.

Therefore, it is verified that the normalized training loss is subject to the chi-square distribution ${\chi ^2}\left( {2M} \right)$. 
\end{lemma}

According to Lemma 1, $2M{\xi _1}/{\cal{L}}_{{\rm{E1}}}$ and $2M{\xi _2}/{\cal{L}}_{{\rm{E2}}}$ are both subject to the chi-square distribution ${\chi ^2}\left( {2M} \right)$. We denote ${\kappa} = 2M$ as the degree of freedom in ${\chi ^2}\left( {2M} \right)$. Then, the PDF of ${\xi _1}$ can be represented as 
\begin{equation}
\label{equ.PDF1}
{p_1}\left( x \right) = {\kappa  \over {{\cal{L}}_{{\rm{E1}}}}}{p_{\chi _\kappa ^2}}\left( {{{\kappa x} \over {{\cal{L}}_{{\rm{E1}}}}}} \right),
\end{equation}
and the CDF of ${\xi _2}$ can be expressed as 
\begin{equation}
\label{equ.CDF2}
{F_2}\left( x \right) = {F_{\chi _\kappa ^2}}\left( {{{\kappa x} \over {{\cal{L}}_{{\rm{E2}}}}}} \right).
\end{equation}
Substituting (\ref{equ.PDF1}) and (\ref{equ.CDF2}) into (\ref{equ.possib1}) gives
\begin{equation}
\label{equ.possibac}
\begin{aligned}
   \varepsilon  &= \int_0^\infty  {{p_1}\left( {{x_1}} \right){F_2}\left( {{x_1}} \right)} d{x_1}  \cr 
  &  = \int_0^\infty  {{\kappa  \over {{\cal{L}}_{{\rm{E1}}}}}{p_{\chi _\kappa ^2}}\left( {{{\kappa {x_1}} \over {{\cal{L}}_{{\rm{E1}}}}}} \right)} {F_{\chi _\kappa ^2}}\left( {{{\kappa {x_1}} \over {{\cal{L}}_{{\rm{E2}}}}}} \right)d{x_1}  \cr 
  & \mathop {{\rm{ }} = \hfill}\limits^{{\varsigma _1} = {{\kappa {x_1}} \over {{\cal{L}}_{{\rm{E1}}}}}} \int_0^\infty  {{p_{\chi _\kappa ^2}}\left( {{\varsigma _1}} \right)} {F_{\chi _\kappa ^2}}\left( {{{{{\cal{L}}_{{\rm{E1}}}}{\varsigma _1}} \over {{\cal{L}}_{{\rm{E2}}}}}} \right)d{\varsigma _1}  \cr 
  &  = \int_0^\infty  {{F_{\chi _\kappa ^2}}\left( {{{{\varsigma _1}} \over {1 + {{{\Delta _{{\cal{L}}_{{\rm{E}}}}}} \over {{\cal{L}}_{{\rm{E1}}}}}}}} \right){p_{\chi _\kappa ^2}}\left( {{\varsigma _1}} \right)} d{\varsigma _1}.
\end{aligned}
\end{equation}

As can be seen from (\ref{equ.possibac}), $\varepsilon$ is determined by $\kappa$, which is  related to the sample size, the MSE difference ${\Delta _{{\cal{L}}_{{\rm{E}}}}}$ and the minimal MSE ${{\cal{L}}_{{\rm{E1}}}}$.  We define $\alpha = {\Delta _{{\cal{L}}_{{\rm{E}}}}}/{{\cal{L}}_{{\rm{E1}}}}$, where $\alpha$ can be regarded as the scaled MSE difference. We use $\alpha$ as the performance metric for the ML-based channel estimation. Then, there are only two variables left, i.e., $\kappa$ and $\alpha$. After the confidence level $1-\varepsilon$ is determined, a clear analytical relation between $\alpha$ and $\kappa $ can be derived. 

It is intuitive that the structure of the learning module, including the input dimension and the category of the learning module (linear or non-linear), influences the learning performance. However, the learning performance indicator $\alpha$ is only determined by the training data size indicator $\kappa $. This is because the analytical relation between $\alpha$ and $\kappa $ is derived based on Assumption 2 and Assumption 3. The two assumptions actually exclude the effects of the structure of the learning module from the analysis result since the two assumptions require that the learning module should be approximately linear and the input dimension should be low. We can see that although the two assumptions limit the applicable scenarios of the analysis, they indeed contribute to obtaining a clear relation between the learning performance and the sample size.

\begin{figure}
\begin{centering}
\includegraphics[scale=0.475]{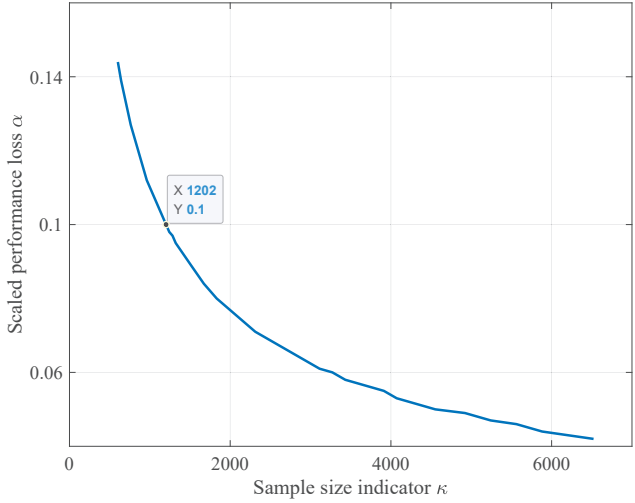}

\end{centering}
\caption{Figure of scaled performance loss upper bound $\alpha$ varied with sample size indicator $\kappa$ when $\varepsilon$ is set to 0.05.}
\label{fig.NumVsPerf}
\end{figure}

The value, 0.95, is usually considered to be an acceptable confidence level \cite{lehmann2006testing}. Therefore, we set $\varepsilon = 0.05$ and plot the curve of $\alpha$ as a function of $\kappa $ in Fig. \ref{fig.NumVsPerf}. The required training data size can be obtained from this curve. As an example, we consider that the learning performance is satisfactory when the MSE difference of the learned estimator is lower than 10\% of the optimal MSE performance. Then, the required training data size corresponds to $\alpha=0.1$ on the curve. It can be seen that when $\kappa $ is above 1200, $\alpha$ is below 0.1. Therefore, it shows that if the learning module is linear and its input dimension is low, we can create a training dataset whose size is only around 600 (the training data size $M=\kappa/2$) and a satisfactory estimator can be learned based on the training dataset. Note that the learning performance indicator $\alpha$ is an upper bound for the scaled MSE difference. The above analysis result may still hold when Assumption 2 and Assumption 3 are slightly violated in practice. In addition, the sufficient dataset size 600 is derived for $\alpha = 0.1$. When other values of the learning performance indicator $\alpha$ are chosen, the sufficient dataset size varies w.r.t $\alpha$.

\section{Performance Evaluation for ML-based Channel Estimation Employing Linear Structure}
\label{sec.EAR}

In the previous section, we present the performance analysis of the ML-based channel estimation and a conclusion is reached that 600 is a sufficient sample size when the linear learning module with a low input dimension is used. To verify this conclusion, we conduct computer simulations to examine the performance of ML-based channel estimation in which the linear learning module is employed. We use stationary channel models in simulations so that the optimal channel estimation has a closed-form and the performance measure, i.e., scaled MSE difference $\alpha$, can be calculated.

\subsection{System Model for Simulation}

We consider an OFDM system with $N$ subcarriers \cite{4267831}. The DC (direct current) carrier and a certain number of carriers at the edges of the spectrum are null, and we denote $K$ as the number of usable subcarriers in an OFDM symbol. CP length ${N_{{\rm{cp}}}}$ is set to $N/4$ and assumed to be over the maximum delay. The channel is assumed to be constant over one OFDM symbol. Time and frequency synchronization are assumed to be accurate as well.

We use an exponentially decaying power-delay profile (PDP), which can be expressed as 
\begin{equation}
\Gamma \left( \tau  \right) = C{e^{ - {\tau  \mathord{\left/
 {\vphantom {\tau  {{\tau _{\max }}}}} \right.
 \kern-\nulldelimiterspace} {{\tau _{\max }}}}}}
\end{equation}
where $C$ is a normalization coefficient and ${{\tau _{\max }}}$ is the maximum delay of the channel.

\begin{figure}
\begin{centering}
\includegraphics[scale=0.8]{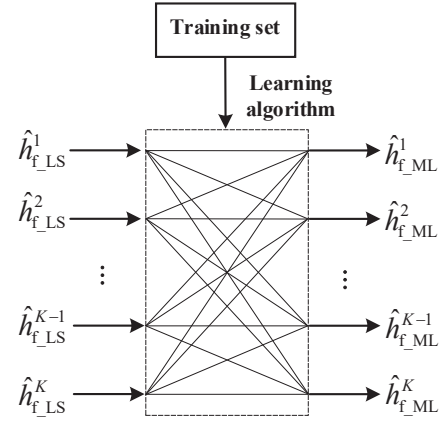}

\end{centering}
\caption{Sketch diagram of ML-based channel estimation employing the linear structure.}
\label{fig.LDDE}
\end{figure}

Referring to the framework in Fig. \ref{fig.MLEstStruct}, the estimator can be interpreted as Fig. \ref{fig.LDDE}. Note that the estimator is not exactly the same as the framework in Fig. \ref{fig.MLEstStruct}. The output of the estimator has the same dimension as the input, while the framework in Fig. \ref{fig.MLEstStruct} only gives a single estimate. However, when it comes to the evaluation of estimation performance, the estimator becomes consistent with the framework. We calculate the average over the MSEs of the channel frequency responses (CFRs) and the output of the estimator is treated as a single estimate like the framework in Fig. \ref{fig.MLEstStruct}. Therefore, the MSE performance of the estimator in Fig. \ref{fig.LDDE} can be used to examine the analysis in Section \ref{sec.PA}.

The estimator in Fig. \ref{fig.LDDE} is actually the realization of (\ref{equ.EstVec}). Since we use a linear learning module and the processing on LS estimates is linear, the function $\bf f$ in (\ref{equ.EstVec}) can be expressed as the multiplication by a coefficient matrix. Therefore, the mathematical expression of the estimator is given by
\begin{equation}
{\bf{\hat h}}_{{\rm{f}}}^{\rm{ML}}={\bf{W}}{{\bf{\hat h}}_{{\rm{f}}}^{\rm{LS}}},
\end{equation}
where ${\bf{W}}$ is a $K \times K$ matrix and contains the parameters of the learning module. ${{{\bf{\hat h}}}_{\rm{f}}^{\rm{LS}}}$ corresponds to ${{\hat {\bf h}}_{\rm{p}}}$ in (\ref{equ.EstVec}) and contains the LS estimates of CFRs, i.e.,
$${\bf{\hat h}}_{\rm{f}}^{{\rm{LS}}}{\rm{ = }}{\left[ {\hat h_{{\rm{f\_LS}}}^1,...,\hat h_{{\rm{f\_LS}}}^K} \right]^{\rm{T}}}. $$
${\bf{\hat h}}_{{\rm{f}}}^{\rm{ML}}$ corresponds to $\hat {\bf h}_{\rm s}$ in  (\ref{equ.EstVec}) and contains the output of the estimator, i.e.,
$${\bf{\hat h}}_{\rm{f}}^{{\rm{ML}}}{\rm{ = }}{\left[ {\hat h_{{\rm{f\_ML}}}^1,...,\hat h_{{\rm{f\_ML}}}^K} \right]^{\rm{T}}}. $$

In Fig. \ref{fig.LDDE}, the training set is $${\cal T}=\left \{ {...,\left( {{{{\bf{\hat h}}}_{\rm{f}}^{\rm{LS}}}\left( m \right),{{\bf{h}}_{\rm{f}}}\left( m \right)} \right),...} \right\},$$ where ${{\bf{h}}_{\rm{f}}} $ contains the actual values of the CFRs, which is the label of training data.

The learning algorithm is to find the ${\bf{W}_ *}$ that minimizes the training loss, which can be formulated as

\begin{equation}
\label{equ.Sepoptimal}
{{\bf{W}}_ * } = \mathop {\arg }\limits_{\bf{W}} \min \sum\limits_m {\left\| {{\bf{W}}{{{\bf{\hat h}}}_{\rm{f}}^{\rm{LS}}}\left( m \right) - {{\bf{h}}_{\rm{f}}}\left( m \right)} \right\|_2^2}.
\end{equation}

The optimization problem of (\ref{equ.Sepoptimal}) has an analytical solution \cite{goodfellow2016deep}
\begin{equation}
\label{equ.optimal}
{{\bf{W}}_*} = {\bf{H}}{\left( {{{{\bf{\hat H}}}^{\rm{H}}}{\bf{\hat H}}} \right)^{ - 1}}{{\bf{\hat H}}^{\rm{H}}},
\end{equation}
where ${\bf{\hat H}} = [ {{{{\bf{\hat h}}}_{\rm{f}}^{\rm{LS}}}\left( 1 \right),...,{{{\bf{\hat h}}}_{\rm{f}}^{\rm{LS}}}\left( M \right)} ]$ is a matrix containing the LS estimation of CFRs in the training set $\cal T$ and ${\bf{H}}$ contains the corresponding true values of CFRs, i.e., ${\bf{H}} = \left[ {{{\bf{h}}_{\rm{f}}}\left( 1 \right),...,{{\bf{h}}_{\rm{f}}}\left( M \right)} \right]$. As we use independent training data, the column vectors in ${\bf{\hat H}}$ are not correlated. Therefore, $( {{{{\bf{\hat H}}}^{\rm{H}}}{\bf{\hat H}}})$ is full rank and invertible.

\subsection{Numerical Results}
\label{Sec.LMLS}

In the simulation of the ML-based channel estimation, we first generate training data to optimize the parameter matrix of the learning module and obtain ${\bf{W}}_ *$. We assume that the correct channel statistics, i.e., the noise variance ${\sigma^2}$ and the channel correlation function, are known. That is, given the dataset size $M$, we can use the true channel model to generate $M$ realizations of channel vectors ${{\bf{h}}_{\rm{f}}}$ and the respective LS estimates ${{{\bf{\hat h}}}_{\rm{f}}^{\rm{LS}}}$. We also assume that the channel realizations are independent to support Assumption 3. After training, we use ${\bf{W}}_ *$ to perform channel estimation. The channel realizations are assumed to be independent of those generated for training so that the estimation performance for whole new data can be shown.

We simulate the LMMSE estimator in (\ref{equ.mmse}) as the optimal estimation $f_{\rm opt}(\cdot)$. The MSE of the LMMSE estimator corresponds to the minimal MSE ${{\cal{L}}_{{\rm{E1}}}}$ and the MSE of the learned estimator corresponds to ${{\cal{L}}_{{\rm{E2}}}}$. Then, the performance measure of ML-based channel estimation $\alpha = {\Delta _{{\cal{L}}_{{\rm{E}}}}}/{{\cal{L}}_{{\rm{E1}}}}$ can be calculated using ${{\cal{L}}_{{\rm{E1}}}}$ and ${{\cal{L}}_{{\rm{E2}}}}$, where ${\Delta _{{\cal{L}}_{{\rm{E}}}}}= {{\cal{L}}_{{\rm{E2}}}} -{{\cal{L}}_{{\rm{E1}}}}$.

\begin{figure}
\begin{centering}
\includegraphics[scale=0.5]{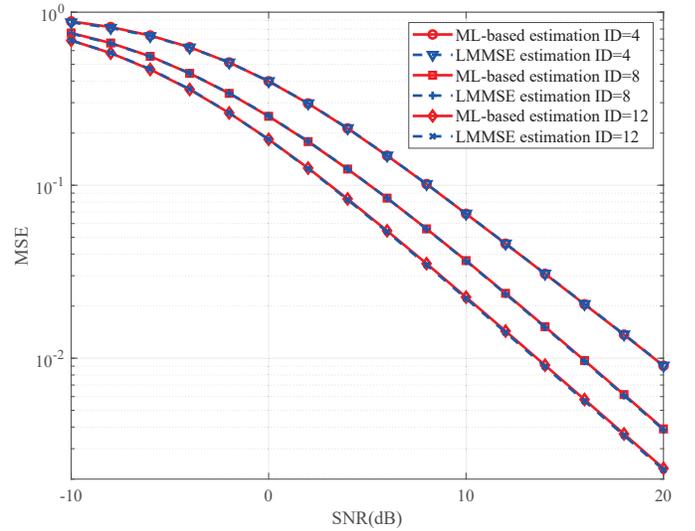}

\end{centering}
\caption{The MSE performance of ML-based channel estimation and LMMSE channel estimation under different SNRs.}
\label{MSE48}
\end{figure}

We first conduct simulation experiments to verify the conclusion reached in Section \ref{Sec.ARSSP} that 600 is a sufficient dataset size when the learning module is linear and its input dimension is low. In our simulations, the input dimension is the number of usable subcarriers per symbol $K$ and we can adjust the number of null subcarriers to control the input dimension. We set the input dimension set as low values including 4, 8, 12. The discrete Fourier transform (DFT) size $N$, the dataset size $M$ and the maximum delay ${{\tau _{\max }}}$ are set to $16$, $600$ and $2$, respectively. Fig. \ref{MSE48} compares the MSE performance of ML-based channel estimation and the LMMSE channel estimation. We can see that the performance of ML-based channel estimation with the three input dimensions is close to that of the LMMSE channel estimation at different signal-to-noise ratios (SNRs). This simulation result verifies that a training data set of size 600 is indeed sufficient for the considered learning module. It also validates our theoretical analysis using hypothesis testing and the derived analytical relation between the dataset size and performance.

\begin{figure}
\begin{centering}
\includegraphics[scale=0.5]{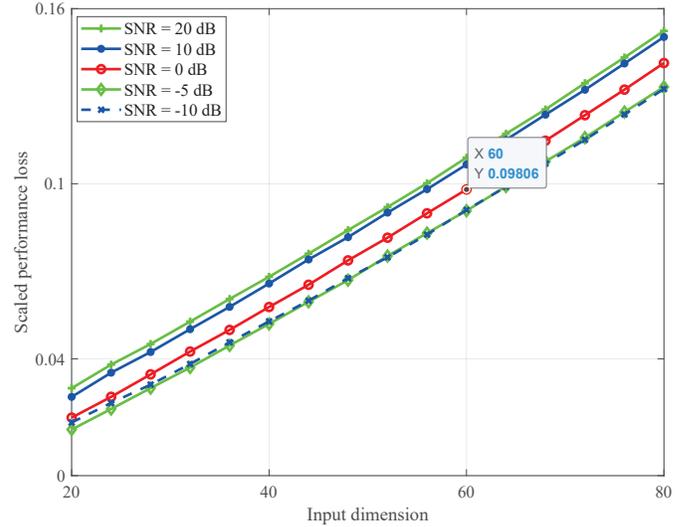}

\end{centering}
\caption{The scaled MSE difference of ML-based channel estimation under different input dimensions.}
\label{AlphaINnum}
\end{figure}

Assumption 3 will not hold with the increasing input dimension as mentioned in Section \ref{Sec.ARSSP}. It can be inferred that the scaled MSE difference $\alpha$ increases as the input dimension increases. Since training data will be insufficient with the growing parameters in the learning module, the learning performance will be degraded. Nevertheless, a training data set of size 600 may still be sufficient when the input dimension is relatively high. As the performance measure $\alpha$ in Section \ref{Sec.ARSSP} is actually the upper bound for the scaled MSE difference, the actual scaled MSE difference may be satisfactory when Assumption 3 is slightly violated. We simulate the scaled MSE difference $\alpha$ of ML-based channel estimation under different values of the input dimension to examine its range. The scaled MSE difference $\alpha$ is presented in Fig. \ref{AlphaINnum} as a function of the input dimension $K$ at SNRs of -10 dB, -5 dB, 0 dB, 10 dB, and 20 dB. It can be observed the curves are very close under high SNRs and low SNRs, respectively, and only have a little difference around the SNR of 0 dB. Furthermore, the scaled MSE difference $\alpha$ is even smaller at lower SNRs, which indicates that the noise variance of the LS estimates does not influence the learning performance. This agrees with the result shown in (\ref{equ.possibac}), which implies that the noise level is not a factor affecting the learning performance. Moreover, the scaled MSE difference $\alpha$ indeed grows with the increasing of the input dimension $K$. As the curve for the SNR of 0 dB has approximately the average performance, we investigate the input dimension range based on this curve. It can be seen that $\alpha < 0.1$ with the input dimension lower than 60. We consider that the learning performance is acceptable when the scaled MSE difference $\alpha < 0.1$. Therefore, the input dimension can increase to 60 with a training dataset of size 600.

\begin{figure}
\begin{centering}
\includegraphics[scale=0.5]{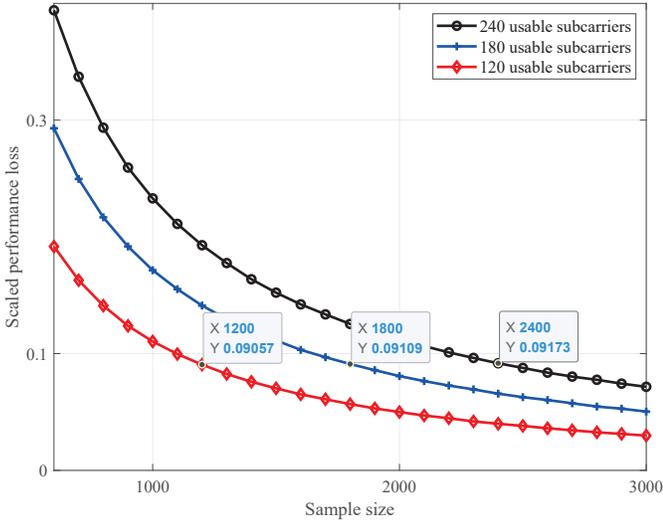}

\end{centering}
\caption{The scaled MSE difference of ML-based channel estimation under different sample sizes.}
\label{AlphaISnum}
\end{figure}

When the input dimension is larger than 60, a training dataset of size 600 is not sufficient and more training data is required. To investigate the required dataset size for ML-based channel estimation with a high input dimension, we set the DFT size $N$ as $256$ and simulate ML-based channel estimation under $K=120, 180, 240$, respectively. In Fig. \ref{AlphaISnum}, we display the scaled MSE difference $\alpha$ with respect to the sample size $M$. As expected, the required sample size grows as the input dimension increases. Furthermore, the required dataset size is approximately in proportion to the input dimension. Based on this result, we can determine the sufficient size of the dataset for high input dimensions. However, further theoretical analysis is required to verify the derived dataset size and it is left as future work.

\section{Performance Evaluation for ML-based Channel Estimation Employing DNN}
\label{sec.CLM}

Deep learning (DL) has attracted much attention recently. Therefore, in this section, we evaluate the performance of ML-based channel estimation employing the deep neural network (DNN) with the derived training dataset size 600. Note that the analysis in Section \ref{Sec.ARSSP} may not hold for this situation. This is because the DNN can approximate a non-linear function, which violates Assumption 2, and commonly has a large number of parameters, which violates Assumption 3.

\subsection{System Model for Simulation}
We simulate the same OFDM system as above but consider the quasi-stationary scenario, where the maximum delay ${{\tau _{\max }}}$ is assumed to be a random variable. As explained in Section \ref{sec.CE}, the closed-form of the optimal channel estimation is hard to derive. However, it is the scenario where ML-based channel estimation shines. The estimator can be directly trained to approximate some complicated expression for the quasi-stationary channel condition since DNN can be regarded as a universal function approximator \cite{8672767}.

\begin{figure}
\begin{centering}
\includegraphics[scale=0.8]{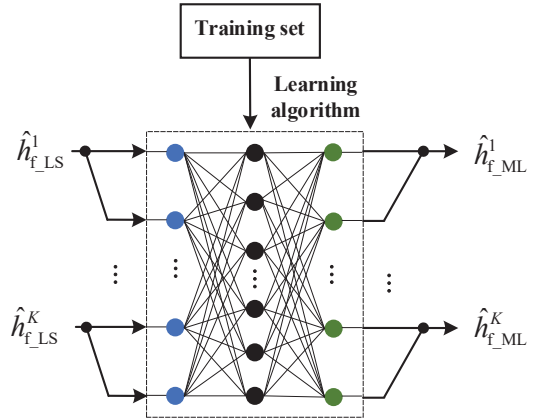}

\end{centering}
\caption{Sketch diagram of ML-based estimation employing the linear structure.}
\label{fig.ANNDDE}
\end{figure}

The structure of the estimator is illustrated in Fig. \ref{fig.ANNDDE}. DNN works as the function $\bf f$ in (\ref{equ.EstVec}), which maps the LS estimates to the improved estimation results. The employed DNN has three layers, including one input layer, one hidden layer, and one output layer. The numbers of neurons in each of the three layers are $2K$, $4K$, $2K$, respectively. Since a DNN cannot get complex numbers as input and also cannot provide them as output, we take the real part and imaginary part of a complex number as two input variables and combine two output variables to generate a complex number. Therefore, the input and output dimensions are twice of the subcarriers' number. The activation function in input and output layers is linear, while the Sigmoid function is used in the hidden layer. 

\subsection{Numerical Results}

The frequency correlation function of channel responses is the Fourier transform of the PDP. Since the maximum delay ${{\tau _{\max }}}$ is a random variable, the channel correlation function is unknown and the LMMSE channel estimation cannot be performed. To cope with the uncertainty of the channel correlation function, the robust estimator proposed in \cite{701317} can be applied. The robust estimator is actually the LMMSE estimation using a uniform PDP based correlation function. The maximum delay of the PDP uses the upper bound of ${{\tau _{\max }}}$, i.e., the possibly largest value of ${{\tau _{\max }}}$. The robust estimator is also called the robust LMMSE estimator in this paper.

\begin{figure}
\begin{centering}
\includegraphics[scale=0.47]{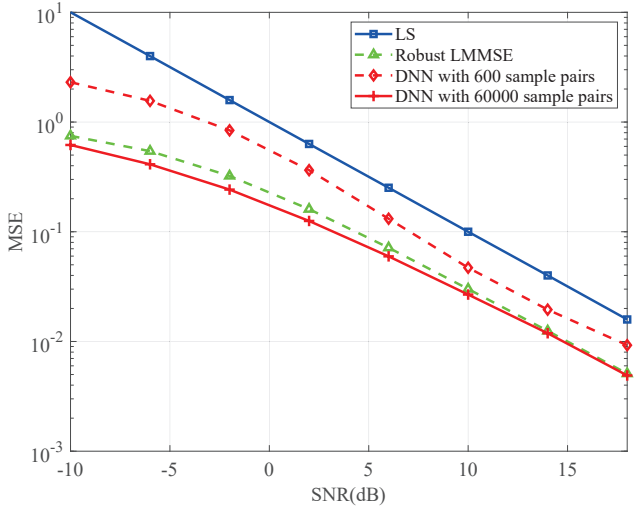}

\end{centering}
\caption{The MSE performance of LS estimator, robust LMMSE estimator, and ML-based estimator under quasi-stationary channel with different SNRs.}
\label{Fig.Unstationay}
\end{figure}

In simulations, the DFT size $N$ is set to $64$ and the valid subcarrier number $K$ is $60$. The maximum delay ${{\tau _{\max }}}$ is drawn from a uniform distribution within $\left[ {1,2,..,16} \right]$. Fig. \ref{Fig.Unstationay} shows the MSE performance of ML-based channel estimation and conventional channel estimation methods including LS estimation and robust LMMSE estimator. When a large dataset is provided for training, the ML-based estimator achieves a significant improvement over the LS estimator and outperforms the robust LMMSE estimator. However, when the sample size is reduced to $600$, huge performance degradation is observed in the ML-based channel estimation. It shows that the derived analytical relation between the training dataset size and performance is not suitable for DL techniques since Assumption 2 and Assumption 3 are violated when DL techniques are used. To derive the sufficient size of the training dataset for ML-based channel estimation employing DL techniques, one may only need to reconsider the distribution of the training loss since the performance analysis using hypothesis testing in Section \ref {SVML} still holds for DL techniques.

\section{ Design of ML-based Channel Estimation with Limited Training Data}
\label{sec.MLCS}

In the previous studies of ML-based channel estimation, it is usually assumed that sufficient training data, e.g., tens of thousands of samples \cite{8272484}, are available. However, in communications systems, generating training data may lead to the reduction of system efficiency. Considering the system efficiency, the available training data may be limited in some applications, e.g., only hundreds of training data. Therefore, we discuss the ML-based channel estimation with a small training dataset in this section.

A linear learning module is preferable for ML-based channel estimation with limited training data compared to complex neural networks. As seen in Fig. \ref{Fig.Unstationay}, although DNN has better potential than the linear structure, it suffers from unacceptable performance degradation when the training data is limited. Similarly, the input dimension should be relatively low since the learning module with a high input dimension requires a large amount of training data as shown in Fig. \ref{AlphaISnum}. Therefore, when the size of the training data is small, the ML-based channel estimation may employ a linear learning module with a low input dimension.

In this case, our theoretical results in Section \ref{Sec.ARSSP} can be used to make quantitative predictions for the learning performance of ML-based channel estimation. Then, the learning module becomes a white box because its performance is predictable given the size of the training dataset. Additionally, our analysis results can help design ML-based channel estimation.

As an example, we consider the design of ML-based channel estimation for an OFDM system with 512 subcarriers, in which 480 are valid, and only 600 sample pairs are available for training. We choose the linear structure as the learning module. As for the input dimension, considering the correlations between subcarriers, the potentially optimal performance may be achieved by setting it as $480$. However, $480$ is too large for the sample size since the provided training data is merely enough for a low input dimension according to our analysis in \ref{Sec.ARSSP}. Therefore, we partition the OFDM symbol into subsymbols consisting of much fewer subcarriers as in \cite{701321} and filter over these subsymbols. In this way, the input dimension can be reduced. We consider that the learning performance is acceptable when $\alpha \le 0.1$. Given that the sample size is $600$, the input dimension can be raised to $60$ as shown in Fig. \ref{AlphaINnum}. Therefore, we expect that 60 is a good choice for the input dimension. When the input dimension is 60, the learning performance is already satisfactory, which means that the performance of the learned estimation is close to that of the optimal estimation. If the input dimension is further reduced, the performance of ML-based channel estimation reduces due to the performance degradation of the optimal channel estimation.

\begin{figure}
\begin{centering}
\includegraphics[scale=0.5]{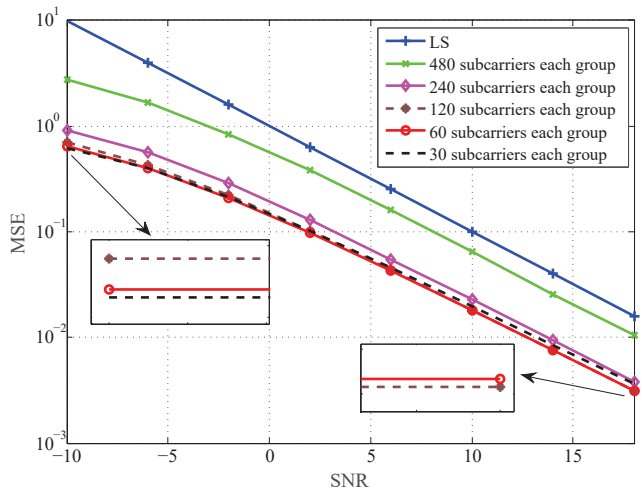}

\end{centering}
\caption{The MSE performance of ML-based channel estimation under different division schemes with different SNRs.}
\label{fig.division}
\end{figure}

We simulate ML-based channel estimation employing the linear structure under several symbol partition schemes. We assume a stationary channel and ${{\tau _{\max }}}$ is set to 64. The results are displayed in Fig. \ref{fig.division}, where the MSE performance of the LS estimator is also plotted as a baseline. Interestingly, we observe that the scheme that partitions the symbols into subsymbols containing 60 subcarriers each, i.e., the input dimension 60, indeed performs the best. When the input dimension is 240 or 480, a significant performance loss is observed due to the lack of training data. The MSEs for the input dimensions 30, 60, and 120 are close and their MSE differences change w.r.t. the SNR. At low SNRs, the input dimension 30 achieves the best performance. This is because the MSE value is large at low SNRs and the MSE of ML-based channel estimation can be reduced significantly by reducing the scaled MSE difference $\alpha$. The ML-based channel estimation with a lower input dimension has smaller $\alpha$ and thus tends to have better performance. However, when the MSE value is quite small at high SNRs, the performance improvement brought by reducing $\alpha$ may be ignorable. Then, the ML-based channel estimation with a high input dimension tends to have better performance since it has a lower achievable MSE. Therefore, the input dimension 120 shows the best performance and the input dimension 30 performs the worst within the three input dimensions at high SNRs. In contrast, the input dimension 60 achieves a good tradeoff between the scaled MSE difference $\alpha$ and the achievable MSE and thus has a moderate performance at all SNRs.

\section{Conclusion}
\label{sec.Conl}

In this paper, we have presented the performance analysis of ML-based channel estimation. The MSE upper bound of ML-based channel estimation is investigated using hypothesis testing. Moreover, we propose the scaled MSE difference, which divides the potentially minimal MSE with the difference between the MSE of ML-based channel estimation and the potentially minimal MSE, as the performance measure. An analytical relation between this performance measure and the sample size is derived for ML-based channel estimation employing the linear learning module with a low input dimension. We plot a curve for this analytical result and derive the size of the training dataset for a given scaled MSE difference $0.1$, which is verified through simulations. Our performance analysis is also validated in the simulation experiments. We discuss the ML-based channel estimation for the scenario where the available training data is limited and apply our analysis results to support the design of the estimator.

Future work may reconsider the statistical model of hypothesis testing and derive the theoretical performance evaluation for ML-based channel estimation employing other learning structures, e.g., the DNN.


%


\section*{Acknowledgment}

This work was supported in part by the National Science Foundation of China (NSFC) (Grants 61931020).

\ifCLASSOPTIONcaptionsoff
  \newpage
\fi



%

\bibliographystyle{IEEEtran}
\bibliography{TCOM-TPS-20-0891}

%
%

%

%

\begin{IEEEbiography}[{\includegraphics[width=1in,height=1.25in,clip,keepaspectratio]{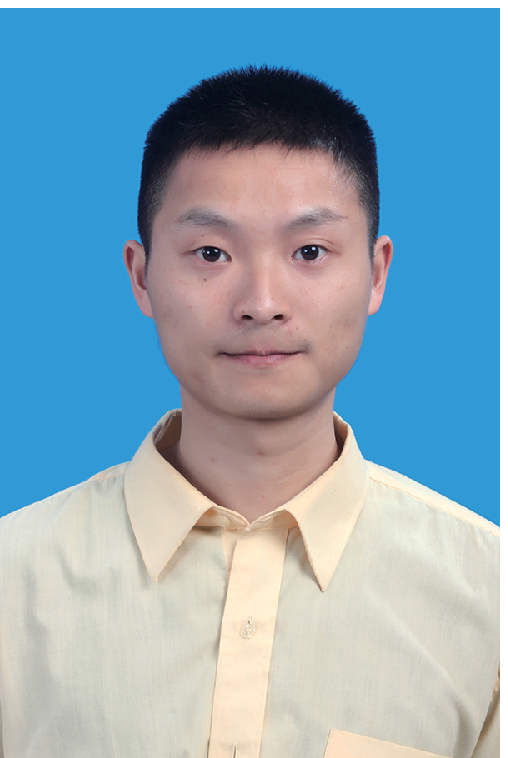}}]{Kai Mei}
received his Master's degree from National University of Defense Technology in 2017. Currently, he is studying at National University of Defense Technology for a Ph.D. degree. He has been a visiting Ph.D. student with the University of Oulu in Finland from 2019 to 2020. His research interests include synchronization and channel estimation in OFDM systems and MIMO-OFDM systems, and machine learning applications in wireless communications.
\end{IEEEbiography}
\begin{IEEEbiography}[{\includegraphics[width=1in,height=1.25in,clip,keepaspectratio]{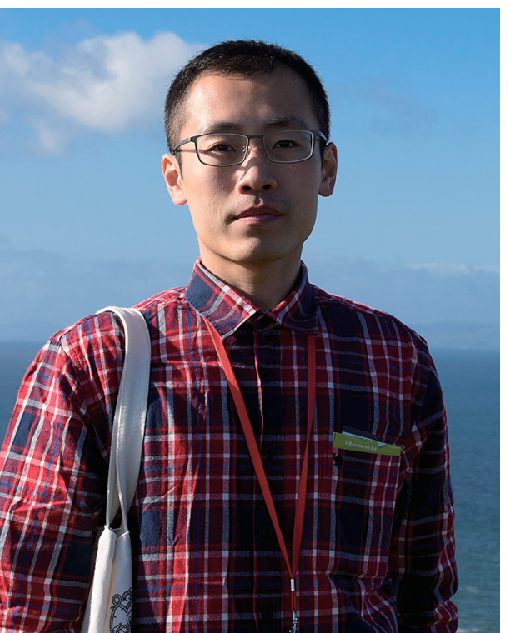}}]{Jun Liu}
received the B.S. degree in optical information science and technology from the South China University of Technology (SCUT), in 2015, and the M.E. degree in communications and information engineering from the National University of Defense Technology (NUDT), Changsha, China, in 2017, where he is currently pursuing the Ph.D. degree with the Department of Cognitive Communications. He is currently a visiting Ph.D. student with the University of Leeds. His current research interests include machine learning with a focus on shallow neural networks applications, signal processing for broadband wireless communication systems, multiple antenna techniques, and wireless channel modeling.
\end{IEEEbiography}
\begin{IEEEbiography}[{\includegraphics[width=1in,height=1.25in,clip,keepaspectratio]{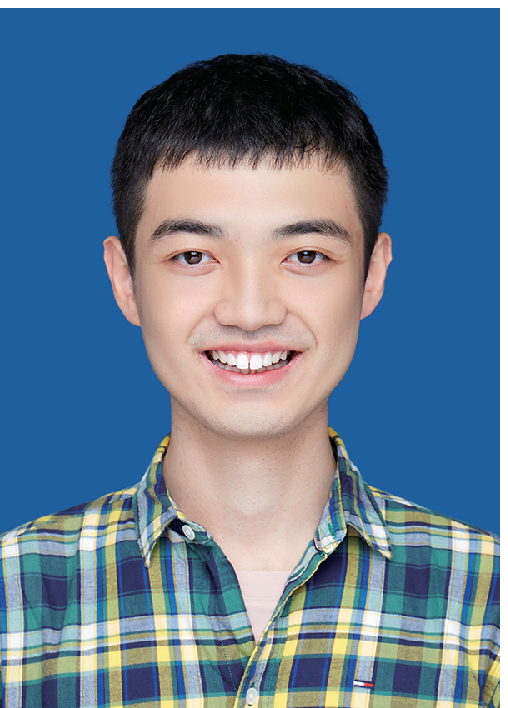}}]{Xiaochen Zhang}
received the B.S. degree from the National University of Defense Technology, Changsha, China, in 2018, where he is currently pursuing the Ph.D. degree with the College of Electronic Science and Engineering. His research interests include resource allocation, multiaccess edge computing, machine learning, and channel modeling.
\end{IEEEbiography}
\begin{IEEEbiography}[{\includegraphics[width=1in,height=1.25in,clip,keepaspectratio]{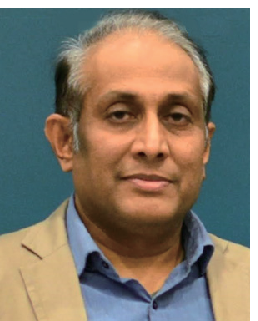}}]{Nandana Rajatheva}
received the B.Sc. (Hons.) degree in electronics and telecommunication engineering from the University of Moratuwa, Sri Lanka, in 1987, ranking first in the graduating class, and the M.Sc. and Ph.D. degrees from the University of Manitoba, Winnipeg, MB, Canada, in 1991 and 1995, respectively. He is currently a Professor with the Centre for Wireless Communications, University of Oulu, Finland. He was a Canadian Commonwealth Scholar during the graduate studies in Manitoba. He held a Professor/Associate Professor positions with the University of Moratuwa and the Asian Institute of Technology, Thailand, from 1995 to 2010. He is currently leading the AI-driven Air Interface design task in Hexa-X EU Project. He has coauthored more than 200 refereed papers published in journals and in conference proceedings. His research interests include physical layer in beyond 5G, machine learning for PHY and MAC, sensing for factory automation and channel coding.
\end{IEEEbiography}
\begin{IEEEbiography}[{\includegraphics[width=1in,height=1.25in,clip,keepaspectratio]{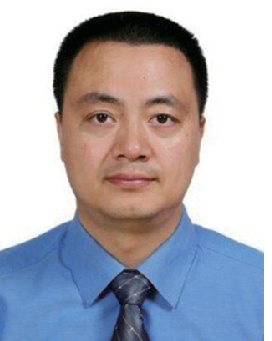}}]{Jibo Wei}
received his BS degree and MS degree from National University of Defense Technology, Changsha, China, in 1989 and 1992, respectively, and the PhD degree from Southeast University, Nanjing, China, in 1998, all in electronic engineering. He is currently a professor of the Department of Communication Engineering of NUDT. His research interests include wireless network protocol and signal processing in communications, cooperative communication, and cognitive network. He is the member of the IEEE Communication Society and also the member of the IEEE VTS. He is a Senior Member of China Institute of Communications and a Senior Member of China Institute of Electronics respectively. He is also an editor of the journal of China Communications.
\end{IEEEbiography}
%
%
%




\end{document}